\documentclass[aps,pre,twocolumn,showpacs]{revtex4}
\usepackage{amsmath}
\usepackage{times}
\usepackage{epsfig}
\usepackage{epstopdf}
 \usepackage{graphicx}

\begin{document}

\title{Mandelbrot Law of Evolving Networks}

\author{Xue-Zao Ren$^{1}$}
%\email{renxuezao@swust.edu.cn}
\author{Zimo Yang$^{2}$}
%\email{yangzimo415@gmail.com}
\author{Bing-Hong Wang$^{1,3}$}
%\email{bhwang@ustc.edu.cn}
\author{Tao Zhou$^{2,3}$}
\email{zhutou@ustc.edu}
\affiliation{
$^1$School of Science, Southwest University of Science and Technology,
 Mianyang 621010, People's Republic of China\\
$^2$Web Sciences Center, University of Electronic Science and
Technology of China, Chengdu 610054, People's Republic of China\\
%$^3$Department of Physics, The Chinese University of Hong Kong,
%Shatin, Hong Kong, People's Republic of China\\
$^3$Department of Modern Physics, University of Science and
Technology of China, Hefei 230026, People's Republic of China
%$^5$Research Center for Complex System Science, University of
%Shanghai for Science and Technology, Shanghai 200093,
%People's Republic of China
 }

\begin{abstract}
Degree distributions of many real networks are known to follow the
Mandelbrot law, which can be considered as an extension of the power
law and is determined by not only the power-law exponent, but also
the shifting coefficient. Although the shifting coefficient highly
affects the shape of distribution, it receives less attention in the
literature and in fact, mainstream analytical method based on
backward or forward difference will lead to considerable deviations
to its value. In this Letter, we show that the degree distribution
of a growing network with linear preferential attachment
approximately follows the Mandelbrot law. We propose an analytical
method based on a recursive formula that can obtain a more accurate
expression of the shifting coefficient. Simulations demonstrate the advantages of our method. This work
provides a possible mechanism leading to the Mandelbrot law of
evolving networks, and refines the mainstream analytical methods for
the shifting coefficient.

\end{abstract}

 \maketitle

%\section{Introduction}

Many systems can be described as networks
\cite{Strogatz:2001,Dorogovtsev:2002,Albert:2002,Newman:2003}, in
which, the nodes correspond to the elements and the links to the
relations between elements. Uncovering the mechanisms underlying the
structural features of real networks is one of the most significant
challenges in network science. Two pioneering models, respectively
for small-world \cite{Watts:1998} and scale-free networks
\cite{Barabasi:1999Sci}, give explanations for many real phenomena,
such as, the logarithmic growth of average distance, the power-law
degree distribution, and the high clustering coefficient.  With the
idea of `rich get richer', the Barab\'{a}si-Albert (BA) network \cite{Barabasi:1999Sci} embodies two mechanisms: growth and preferential
attachment. That is, at each time step, a new node is added and
connected to a few old nodes with probability proportional to their
degree as:
\begin{equation}
\Pi(k_{i})=k_{i}/\sum_{j}k_{j},
\end{equation}
where $k_i$ is the degree of node $i$, and $j$ runs over all old
nodes. The analytical solution of the degree distribution,
\begin{equation}
p(k)=2m^{2}k^{-3},
\end{equation}
can be obtained by applying the mean-field approximation
\cite{Barabasi:1999Sci,Barabasi:1999PhyA}, in which $2m$ is the
average degree of the network.

Unfortunately, for many real networks, the degree distributions are
different from exactly power laws \cite{Newman:2005,Clauset:2009}.
For example, the scientific collaboration networks can be better
characterized by the power-law distributions with exponential cutoff
\cite{Newman:2001}, the degree distributions of the email networks
\cite{Newman2002}, some collaboration networks \cite{Zhang2006}, and
online user-object bipartite networks \cite{Shang:2010} obey the
stretched exponential forms, and the double power-law distribution
seems a better way to describe the air transportation networks
\cite{Li2004,Liu:2007,Bagler2008}. In this Letter, we focus on the
\emph{Mandelbrot law} or called the \emph{shifted power law}
\cite{Mandelbrot:1965}, which can be written as:
\begin{equation}
p(k)\propto(k+c)^{-\gamma},
\end{equation}
where $\gamma$ is the power-law exponent and $c$ is the shifting
coefficient. In fact, for the well-known BA model, the
degree distribution
\begin{equation}
p(k)=\frac{2m(m+1)}{(k+2)(k+1)k} \approx 2m^{2}k^{-3},
\end{equation}
obtained by the master equation \cite{Dorogovtsev:2000}, is not an
exactly power-law distribution. This distribution can be
approximated as $p(k)\propto(k+1)^{-3}$, which also satisfies the
Mandelbrot law with $\gamma=3$ and $c=1$.

\begin{table}
\caption{Fitting exponents and errors for power law and Mandelbrot law on six real networks. P-error and M-error stand for square errors from power-law fitting and Mandelbrot-law fitting, respectively. The smaller the error, the better the fitting.}
\begin{center}
\begin{tabular} {cccccc}
  \hline \hline
   Networks    & $\gamma$($c=0$)  &  P-error & $\gamma$ & $c$ & M-error  \\
   \hline
   SC-Small & 2.4 & 5.69 &	4.3 & 10.5	& 1.81 \\
   SC-Large & 2.3 & 6.38 &	3.6 & 10.3	& 2.00 \\
   PPI & 3.0 & 1.94 & 2.9 & -0.1 & 1.90 \\
   Slashdot & 1.7 & 3.99 & 1.9 & 1.2 & 3.33 \\
   USAir & 1.6 & 3.56 &	1.7 & -0.9 & 3.29 \\
   UCI & 1.8 & 3.74 & 1.9 & 0.4 & 3.63 \\
   \hline \hline
    \end{tabular}
\end{center}
\end{table}

\begin{figure}
\scalebox{0.145}[0.145]{\includegraphics{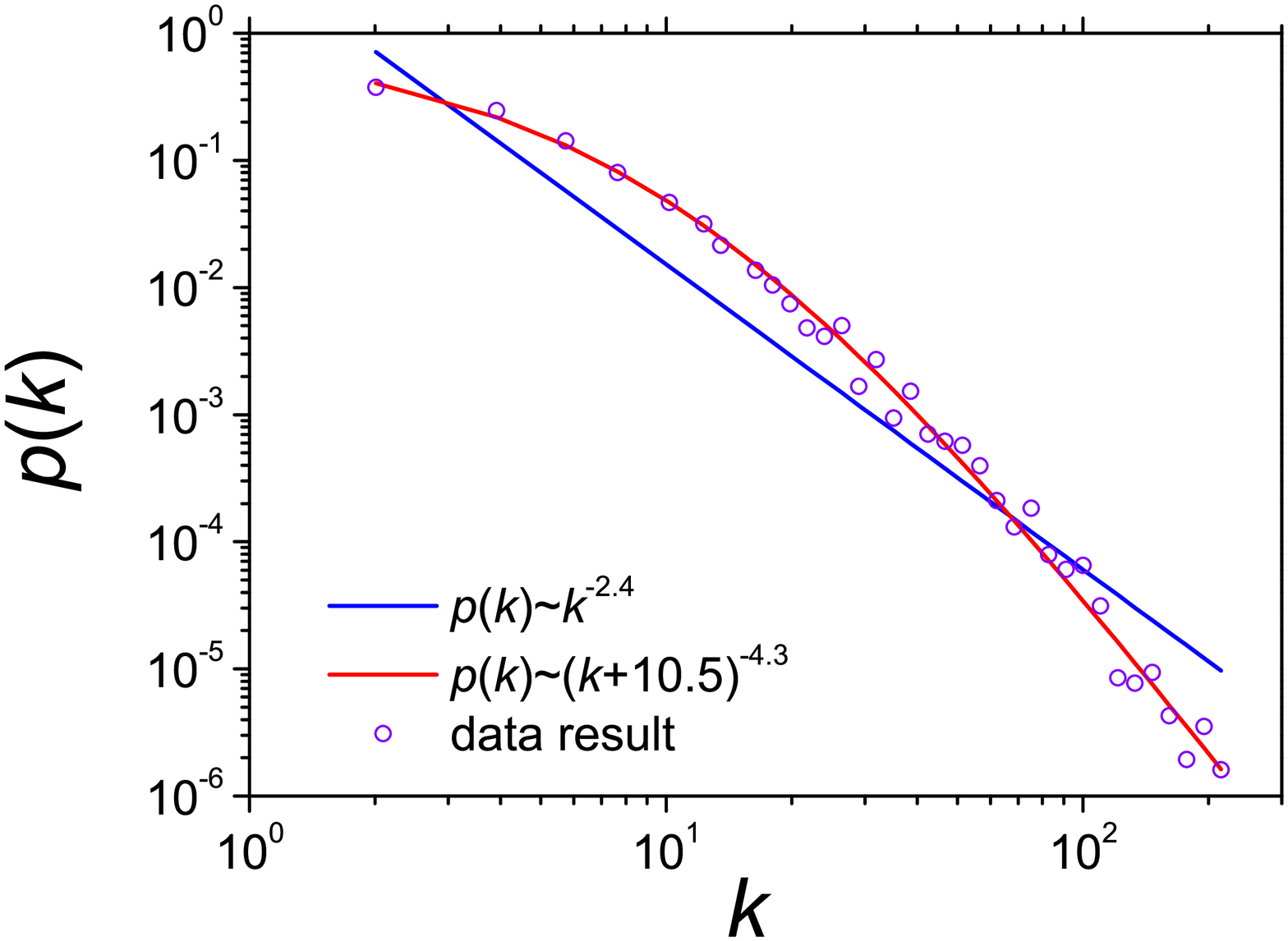}}
\scalebox{0.145}[0.145]{\includegraphics{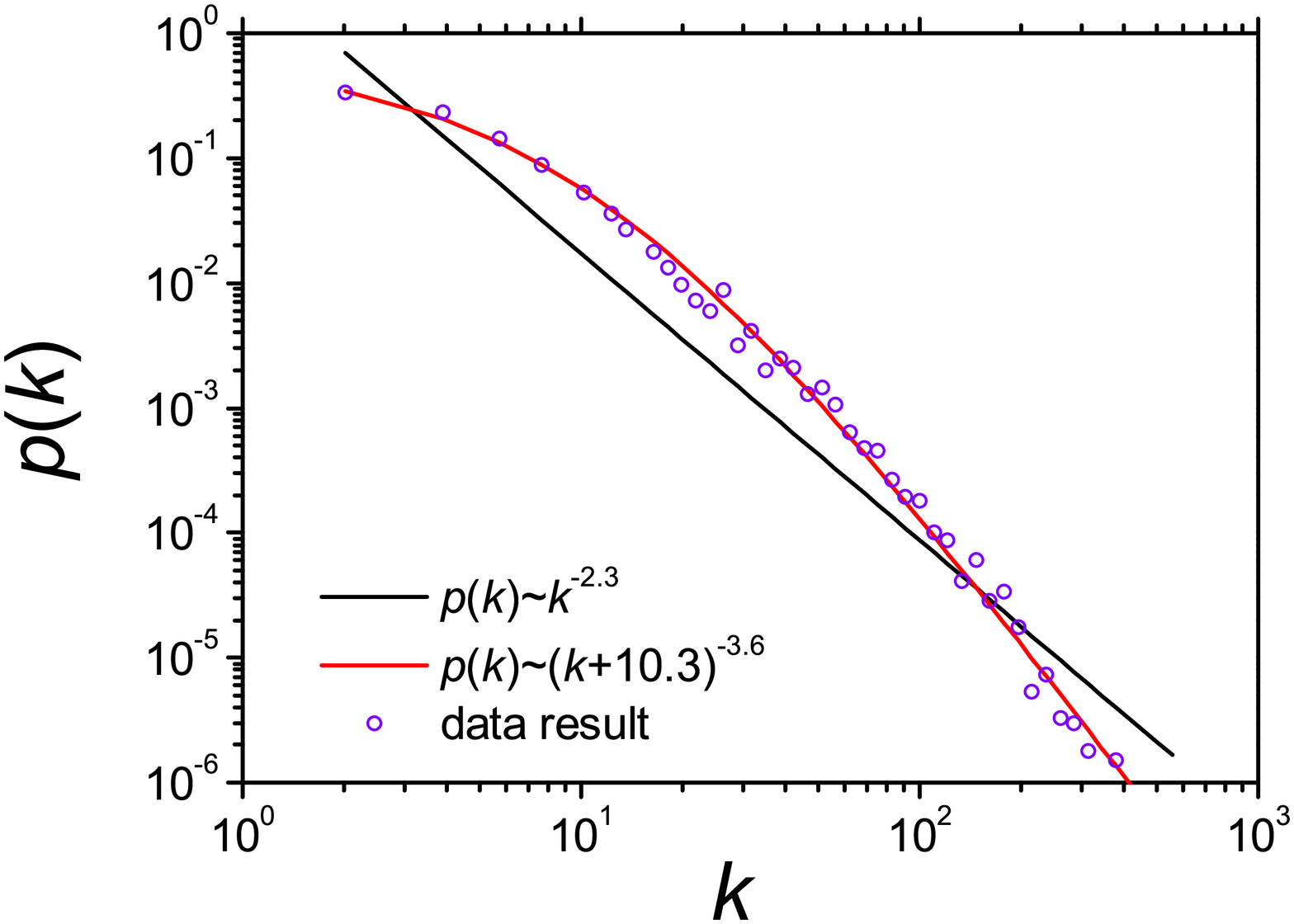}}
\caption{Illustration of power-law and Mandelbrot-law fittings for two scientific collaboration networks. The left and right plots respectively display the networks of smaller and larger sizes (i.e., SC-Small and SC-Large). Obviously, the Mandelbrot law performs much better.}
\end{figure}

Recently, the Mandelbrot law has been applied to
characterize the degree distributions of some real networks
\cite{Chang:2007,Wang:2009}. Here we test the validity of the Mandelbrot law on six real networks: (i) \emph{SC-Small}.--A scientific collaboration network according to e-print manuscripts on condense matter physics from 1995 to 1999 in arxiv.org \cite{Newman:2001}; (ii) \emph{SC-Large}.--Similar to SC-Small but based on manuscripts published from 1993 to 2003 \cite{Leskovec2007}; (iii) \emph{PPI}.--A protein-protein interaction network of yeast \cite{Jeong2001};  (iv) \emph{Slashdot}.--A social network consisted of friend/foe links, attracted from the online service website, Slashdot \cite{Kunegis2009}. (v) \emph{USAir}.--An air transportation network in the United States \cite{Colizza2007}. (vi) \emph{UCI}.--A social network of students at University of California, Irvine \cite{Opsahl2009}. We try the least square method for both power law and Mandelbrot law, and the fitting exponents as well as square errors are shown in Table 1. From this table, we could conclude that: (i) Using the Mandelbrot law can generally improve the fitting accuracy compared with the power law since the former has one more coefficient; (ii) Sometimes the two fitting methods give more or less the same errors, and in these cases, the two power-law exponents are almost the same and the shifting coefficient is usually close to zero; (iii) Sometimes applying the Mandelbrot law can largely improve the fitting accuracy, and then the two power-law exponents are far different while the shifting coefficient is much larger than zero and its significant role cannot be neglected. Figure 1 displays the two cases with remarkable differences between two fitting methods, from which the advantage of the Mandelbrot law is demonstrated.

A number of tools have been developed to get the analytical
solutions of network degree distributions, including the mean-field
approximation, the master equation, the rate equation, and so on
\cite{Barabasi:1999PhyA,Dorogovtsev:2000,Krapivaky:2000,Zhou:2005,Moore:2006}.
Most of these known analytical methods only concentrate on the
power-law exponent, yet pay less attention to the value of shifting
coefficient, which, however, plays a significant role in determining
the shape of degree distributions (see figure 1). Even worse, we will show later
that the widely used difference approximation, no matter forward
difference or backward difference, will result in considerable
deviations to the real value of the shifting coefficient.

We here investigate a model embodying a linear preferential attachment, which
can be considered as an extension of the famous BA model. Initially,
our model starts with a fully connected network with $m_0$ nodes and
$m_0(m_0-1)/2$ links. If the final network size is $S$, it should satisfy
the condition $m_0\ll S$. After initialization, at each time step, a
new node will be added into the network, which will connect to $m$
old nodes. The probability of an old node $i$ to be connected is
linearly correlated with its degree $k_i$, say
\begin{equation}
\Pi(k_{i})=\frac{1}{N}(\alpha k_{i}+\beta),
\end{equation}
where $\alpha$ and $\beta$ are two parameters, and $N$ is the number of nodes at that time step. The self-loop and
multiple links are not allowed. This model will degenerate to the BA
model if $\beta=0$. The parameters $\alpha$ and $\beta$ satisfy the
normalization condition
\begin{equation}
\sum_{k}\pi(k)p(k)=1,
\end{equation}
where $\pi(k)=N\Pi(k)$ and $\pi(k)p(k)$ is the probability a selected old node is of degree $k$.
That is
\begin{equation}
\sum_{k}\pi(k)p(k)=\sum_{k}(\alpha k+\beta)p(k)=2m\alpha+\beta=1.
\end{equation}
It is equivalent to:
\begin{equation}
\alpha=\frac{1}{2m}(1-\beta).
\end{equation}

\begin{figure}
\scalebox{1.2}[1.2]{\includegraphics{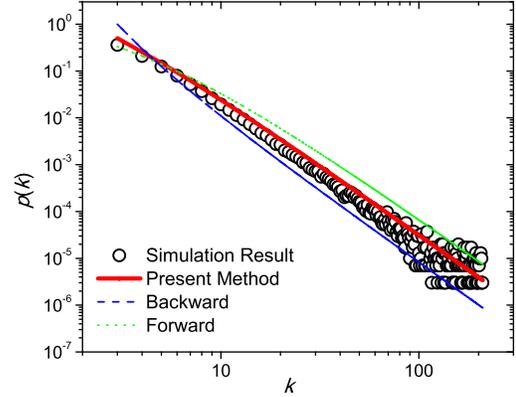}} \caption{Degree
distribution of the modeled network with $\beta=0$ and $m=3$. Since
$\beta=0$, it is equivalent to a BA network. Compared with the
results of backward difference approximation (blue dash line) and
forward difference approximation (green dot line), the present
solution (red solid line) is closer to the simulation result (round
donuts). The network size is $S=10^4$ and the results are obtained
by averaging over 100 independent realizations.}
\end{figure}

%\section{Analysis}

The rate equation is based on the assumption that the added nodes
and links, during a time step, have no influence on the
degree distribution of the network, namely in the thermodynamic
limit, the degree distribution approaches to a steady
form. Denoting $p(k)$ the steady degree distribution and $N$ the
number of nodes in the current time step, if $N$ is large enough,
then the number of nodes with degree $k$ is approximated to $Np(k)$.
Analogously, the number of nodes with degree $k$ in the next step is
$(N+1)p(k)$. With $m$ links added, the number of nodes with degree $k$ in the time step $N+1$ reads
\begin{equation}
(N+1)p(k)=Np(k)+m\pi(k-1)p(k-1)-m\pi(k)p(k)+\delta_{km},
\end{equation}
where $m\pi(k-1)p(k-1)$ and $m\pi(k)p(k)$ represent, respectively,
the number of nodes whose degree changes from $k-1$ to $k$ and from $k$ to
$k+1$ in this time step. $\delta_{km}$ accounts for the specific
degree equal to $m$, namely $\delta_{km}=1$ when $k=m$ and
$\delta_{km}=0$ otherwise. Eq. (9) is the familiar form of the
well-known rate equation \cite{Krapivaky:2000,Zhou:2005}, which is usually solved by the difference approximation, however,
here we will show a different analytical method, and will later
compare our results with the ones obtained by the difference
approximation.

Eq. (9) is equivalent to
\begin{equation}
    \begin{cases}
    p(m)=\frac{1}{1+m\pi(m)}, k=m, \\
    p(k)\left[1+m\pi(k)\right]=m\pi(k-1)p(k-1), k>m. \\
    \end{cases}
\end{equation}
Reminding the linear relation
\begin{equation}
\pi(k)=\alpha k+\beta,
\end{equation}
considering Eq. (8), the probability of a newly added link
connecting to an old node with minimum degree is
\begin{equation}
\pi(m)=\frac{1-\beta}{2m}m+\beta=\frac{1+\beta}{2}.
\end{equation}
Clearly, this probability should be no less than zero and no larger
than one, and thus $-1\leq\beta\leq1$. According to Eq. (10), the
probability density of $m$-degree nodes is
\begin{equation}
p(m)=\frac{1}{1+m\pi(m)}=\frac{2}{2+m(1+\beta)}.
\end{equation}
Substituting Eq. (8) into Eq. (10), we get
\begin{equation}
p(k)\left[k+\frac{2(1+m\beta)}{1-\beta}\right]=\left[k+\frac{2m\beta}{1-\beta}-1\right]p(k-1).
\end{equation}
Specifying:
\begin{equation}
a\equiv\frac{2m\beta}{1-\beta}-1, \texttt{   } b\equiv\frac{2(1+m\beta)}{1-\beta},
\end{equation}
then Eq. (14) can be rewritten in a recursive formula as
\begin{equation}
p(k)=\frac{k+a}{k+b}p(k-1).
\end{equation}
Taking logarithm in both sides of Eq. (16), we get
\begin{equation}
\log\frac{p(k)}{p(k-1)}=\log\frac{k+a}{k+b}.
\end{equation}
With the ansatz that $p(k)$ follows the Mandelbrot law, substituting
Eq. (3) into Eq. (17), we obtain the relationship between
power-law exponent $\gamma$ and shifting coefficient $c$ as
\begin{equation}
\log\frac{k+a}{k+b}=\gamma \log\frac{k-1+c}{k+c},
\end{equation}
which is equivalent to:
\begin{equation}
\log\frac{1+a\frac{1}{k}}{1+b\frac{1}{k}}=\gamma\log\frac{1+(c-1)\frac{1}{k}}{1+c\frac{1}{k}}.
\end{equation}
Under the approximation with large $k$, through the second order
Taylor expansion of Eq. (19) with $1/k$ being the variable, we can
get the power-law exponent
\begin{equation}
\gamma=b-a=1+\frac{2}{1-\beta},
\end{equation}
and the shifting coefficient
\begin{equation}
c=\frac{b+a+1}{2}=\frac{1+2m\beta}{1-\beta}.
\end{equation}

\begin{figure}
\scalebox{1.2}[1.2]{\includegraphics{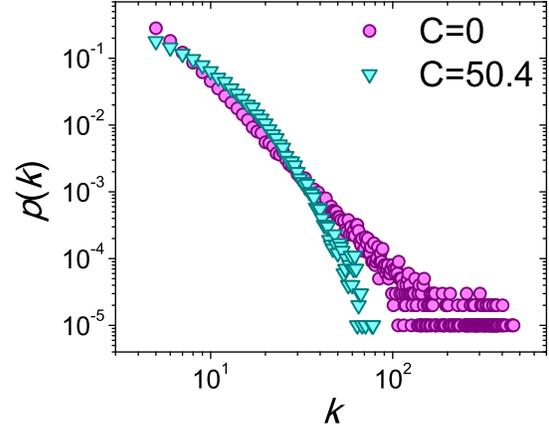}} \caption{The
comparison of degree distributions with different shifting
coefficients given $m=5$ and $S=10000$. Compared with the case of
$\beta=0.8$ and shifting coefficient $c=50.4$ (green
down-triangles), the degree distribution of the none-shifting case
with $\beta=-0.1$ and $c=0$ (purple circles) is much closer to a
straight line in the log-log coordinates. The results are obtained
by averaging over 100 independent realizations.}
\end{figure}

Eq. (20) and Eq. (21) declare that $\gamma$
only depends on $\beta$, while $c$ is related to both $\beta$ and $m$. When $m\beta$ is
very large or $\beta\rightarrow 1$, $a$ and $b$ are both very large,
and the Taylor expansion cannot be applied on Eq. (19). Under such
condition, Eq. (16) can be approximately rewritten as
\begin{equation}
p(k)\approx \frac{a}{b}p(k-1),
\end{equation}
then the degree distribution is close to an exponential
form. It is easy to be understood since when
$\beta\rightarrow 1$, the selection of old nodes is almost random.
When $\beta=0$, $\alpha=\frac{1}{2m}$, our model degenerates to the
BA model, and we get $a=-1$, $b=2$, $\gamma=3$ and $c=1$, with
degree distribution being
\begin{equation}
p(k)=-\frac{2}{\psi(2,m+1)}(k+1)^{-3},
\end{equation}
where
\begin{equation}
\psi(x)=\Gamma'(x)/\Gamma(x)
\end{equation}
is the \emph{Digamma function} with
\begin{equation}
\Gamma(x)=\int_0^\infty e^{-t}t^{x-1}dt
\end{equation}
being the \emph{Gamma function} and
\begin{equation}
\psi(n,x)=\frac{d^{n}\psi(x)}{dx^{n}}.
\end{equation}

We next compare the present method with
methods based on the difference approximation. We first introduce the
backward difference approximation, which assumes
\begin{equation}
\frac{dp}{dk}=p(k)-p(k-1).
\end{equation}
Substituting Eq. (27) into Eq. (16), we get
\begin{equation}
p(k)=\frac{k+a}{k+b}\left[p(k)-\frac{dp}{dk}\right],
\end{equation}
which is equivalent to
\begin{equation}
\frac{dp}{dk}=\frac{a-b}{k+a}p(k)
\end{equation}
that leads to the solution
\begin{equation}
p(k)\propto (k+a)^{-(b-a)}.
\end{equation}
Similarly, the forward difference approximation
assumes
\begin{equation}
\frac{dp}{dk}=p(k+1)-p(k),
\end{equation}
and then Eq. (16) can be rewritten as
\begin{equation}
p(k+1)=\frac{k+a+1}{k+b+1}p(k),
\end{equation}
which is equivalent to
\begin{equation}
\frac{dp}{dk}=\frac{a-b}{k+1+b}p(k).
\end{equation}
In this case, the solution is
\begin{equation}
p(k)\propto (k+b+1)^{-(b-a)}.
\end{equation}

The three methods all indicate that the Mandelbrot law will emerge
from an evolving network with linear preferential attachment, and
give the same power-law exponent $\gamma=b-a$. In contrast,
the shifting coefficient are different:
$c^{\texttt{present}}=\frac{a+b+1}{2}$, $c^{\texttt{backward}}=a$
and $c^{\texttt{forward}}=b+1$. In Fig. 1, we compare the
degree distributions obtained by these three methods with the
simulation results and show that the present method is
more accurate.

Although we usually refer to the concept of scale-free networks,
neither the BA networks nor most real networks have very precisely
power-law degree distributions. The present method suggests that we
can obtain a more precisely power-law distribution by setting a
right $\beta$ that corresponds to a zero shifting coefficient. Since
the degree distribution is
\begin{equation}
p(k)\propto \left(k+\frac{1+2m\beta}{1-\beta}\right)^{1+\frac{2}{1-\beta}},
\end{equation}
the non-shifted degree distribution asks for
\begin{equation}
c=\frac{1+2m\beta}{1-\beta}=0,
\end{equation}
namely $\beta=-\frac{1}{2m}$ and $p(k)\propto k^{3-\frac{2}{2m+1}}$.
Therefore, given the linear preferential attachment, the non-shifted
power-law exponent is determined by the average degree and
can never exceed 3. Figure 2 compares two degree distributions,
respectively with $c=0$ and $c=50.4$, from which one can confirm
that the non-shifted power law is indeed much closer to a straight
line in the log-log coordinates, and the shifting coefficient
largely affects the shape of degree distribution.

In summary, we extend the BA model to an evolving model with
linear preferential attachment and show that the corresponding degree distribution obeys the Mandelbrot law. The shifting coefficient, usually being ignored in the literature, largely affects the shape of degree distribution. In puzzlement, the
backward and forward difference approximations will lead to
different solutions on shifting coefficient. Our analysis indicate
that both of them are inaccurate, and we propose an analytical
method that results in a more accurate solution.

\end{document}